\documentclass[aps,pra,reprint,superscriptaddress,twocolumn,showkeys,showpacs,amsmath,amssymb,longbibliography,floatfix]{revtex4-2}
\usepackage{amsmath,amssymb,bbm,mathrsfs,bm,braket,color,graphicx,comment,amsfonts,dsfont,relsize}
\usepackage[colorlinks,linkcolor=blue,citecolor=blue,urlcolor=blue]{hyperref}
\usepackage[mathscr]{euscript}
\usepackage{ulem}
\usepackage{xcolor}
\usepackage{scalerel}
\usepackage[version=3]{mhchem}

\mathchardef\mhyphen="2D
\newcommand{\Tau}{\mathrm{T}}
\newcommand{\timebraket}[3][]{\left\langle \hspace{#1}\left\langle\vphantom{#2 #3}\right. #2 \left\vert\vphantom{#2 #3} \right.#3\left. \vphantom{#2 #3}\right\rangle \hspace{#1}\right\rangle }

\begin{document}

\title{Floquet Engineering of a Diatomic Molecule Through a Bichromatic Radiation Field}

\author{Edgar Barriga}
\affiliation{Departamento de F\'{\i}sica, Facultad de Ciencias, Universidad de Chile, Santiago, Chile}
\author{Luis E. F. {Foa Torres}}
\email{luis.foatorres@uchile.cl}
\affiliation{Departamento de F\'{\i}sica, Facultad de Ciencias F\'{\i}sicas y Matem\'aticas, Universidad de Chile, Santiago, Chile}
\author{Carlos Cárdenas}
\email{cardena@uchile.cl}
\affiliation{Departamento de F\'{\i}sica, Facultad de Ciencias, Universidad de Chile, Santiago, Chile}
\affiliation{Centro para el Desarrollo de la Nanociencia y la Nanotecnología (CEDENNA), Santiago, Chile}

\begin{abstract}
We report on a theoretical study of a \ce{Cs_2} molecule illuminated by two lasers and show how it can result in novel quantum dynamics. We reveal that these interactions facilitate the bypass of the non-crossing rule, forming Light-Induced Conical Intersections and modifiable avoided crossings. Our findings show how laser field orientation and strength, along with initial phase differences, can control molecular state transitions, especially on the \textit{micromotion} scale. We also discuss extensively how the interaction of radiation with matter gives rise to the emergence of potential energy surfaces of hybrids of radiation and molecular states.  This research advances a technique for manipulating photoassociation processes in \ce{Cs_2} molecules, offering potential new avenues in quantum control.

\end{abstract}

\maketitle

\section{Introduction}

From the inception of quantum theory~\cite{smekal_zur_1923}, light-matter interaction has been a cornerstone of research, leading to innovations like Raman spectroscopy already in 1928, long before the first lasers became available~\cite{raman_negative_1928}, and modern quantum information processing~\cite{konishi_universal_2021}. While earlier studies primarily used light to sense matter properties, recent research is pivoting towards manipulating these properties. This shift leverages the emergence of hybrid states when matter interacts with radiation, possessing unique properties distinct from both matter and light~\cite{oka_floquet_2019} (see also section 11 of~\cite{giustino_2021_2020} for more insights). In recent years, Floquet engineering has gained momentum~\cite{noauthor_quantum_2020}, spawning research in areas like Floquet topological insulators~\cite{rudner_band_2020}, the detection of Floquet-Bloch states~\cite{wang_observation_2013,mahmood_selective_2016}, and the laser-induced Hall effect~\cite{mciver_light-induced_2020}, to name a few.

Beyond solid-state systems, intense laser illumination of molecules may also bring new surprises, while unexpectedly bridging areas. One notable case is the topological frequency conversion reported in ~\cite{martin_topological_2017}, where the illumination of a molecule with two lasers produces an effective two-dimensional topological lattice (with each laser frequency adding an effective dimension)~\cite{crowley_topological_2019}. Another example of interest, and which is the focal point here, concerns light-induced degeneracies (LIDs) in the potential energy surfaces (PES) of molecules and the quest for achieving light-assisted control of those surfaces in general. Questions in this field abound: Can we obtain new features unique to laser-illuminated systems? Can we achieve further control by using a multi-tone driving and in such case which new control paths open-up?

Here we explore the impact of bichromatic illumination on the diatomic molecule \ce{Cs_2}. Leveraging Many-Mode Floquet Theory (MMFT)~\cite{ho1983,ho1984}, and focusing on a model that encompasses the ground state and two excited states of \ce{Cs_2}, we discern three pivotal findings:
\begin{enumerate}
\item[(\textit{i})] We identify diverse types of light-induced avoided crossings. These can be modulated by adjusting the laser frequency and intensity. Notably, the dual-tone driving  provides a mean for a high degree of engineering on the shape of the  light-induced potential energy surfaces of the atoms.

\item[(\textit{ii})] We further probe the influence of the lasers' initial phase on the transition probabilities between molecular states, attributing significant effects to \textit{micromotion}. Simulations indicate that \textit{micromotion} becomes crucial at laser intensities around $10^{11}$~W/cm$^2$ and above. 
\item[(\textit{iii})] By combining \textit{i} and \textit{ii}, we demonstrate the potential of laser-assisted processes to amplify photoassociation of Cs$_2$.
\end{enumerate}

We selected \ce{Cs_2} as our diatomic molecule of interest due to its rich theoretical and experimental research history~\cite{fioretti2000}. Its formation, through the association of ultra-cold Cs atoms, serves as an archetype of ultra-cold chemistry [refs]. However, our insights extend beyond \ce{Cs_2} and can be applied to other diatomic or even polyatomic molecules, especially when a single nuclear degree of freedom (e.g., bond, angle, dihedral angle) is paramount.

The subsequent sections will introduce our theoretical framework, followed by a detailed discussion of our results.

\section{Theoretical framework}\label{sec:equations}
The time-dependent Schr\"odinger equation for a molecular system interacting with a two-color laser field has the form
\begin{align}\label{eq:schrodinger}
  i\hbar\frac{\partial}{\partial t}\Ket{\Psi(t)} &= {\cal H}(t)\Ket{\Psi(t)}\,,\\
  \nonumber {\cal H}(t) &= \hat T+{\cal H}_e+\mathcal{H}_{\text{int}}(t)\,,
\end{align}
where $\hat T$ is the nuclear kinetic energy operator, ${\cal H}_e$ stands for the electronic Hamiltonian and $\mathcal{H}_{\text{int}}(t)$ is the time-dependent interaction term that couples the electronic motion with the field, which for continuous wave lasers in the length gauge and the dipole approximation is given by 
\begin{align}\label{eq:H_int}
  \mathcal{H}_{\text{int}}(t) &= E_{1}e^{-i\omega_1 t}+E_{2}e^{-i\omega_2 t}+\rm{c.c.}\,,
\end{align}
with
\begin{align}\label{eq:polarization}
  E_{k} &= 
  \begin{cases}
    \displaystyle \frac{1}{2}\epsilon_0^{(k)}e^{-i\delta_k}\sum_{n=1}^{N_e} R_n \cdot \hat u_k\quad &\substack{\text{Linearly } \\ \text{polarized}}\\
    \displaystyle \frac{1}{2}\epsilon_0^{(k)}\sum_{n=1}^{N_e} R_n\cdot \big(\hat u_k+ip\hat v_k \big) \quad &\substack{\text{Circularly } \\ \text{polarized}} 
  \end{cases}
\end{align}
where $\epsilon_0^{(k)}$ is the field strength, $\delta_k$ the initial phase of the field, $N_e$ is the number of electrons, $\hat u_k$ ($\hat v_k$) is the polarization axis, $p\ (=\pm 1)$ is the handedness of the circular polarization. 

\subsection{Floquet Theory}\label{sec:Floquet}

Since \ce{Cs_2} has large mass ($4.8\times10^5\ m_e$), we neglect the kinetic energy operator and solve the time-dependent Schr\"odinger equation for the electronic wave function with clamped nuclei
\begin{align}\label{eq:schrodinger_electronic}
  i\hbar\frac{\partial }{\partial t}\Ket{\Phi(t)} &= \left({\cal H}_e+{\cal H}_{\text{int}}(t)\right)\Ket{\Phi(t)}.
\end{align}

In our work we are interested in a quasiperiodic bichromatic driving, i.e., $\omega_1/\omega_2$ irrational. The interaction term $\mathcal{H}_{{\rm int}}(t)$ is $2\pi$-periodic in each frequency and therefore the solutions of \eqref{eq:schrodinger_electronic} have the same canonical form~\cite{murdock1978} as the solutions for the periodic case~\cite{shirley1965},
\begin{align}\label{eq:floquet_solutions}
  \Ket{\Phi_{j}(t)} &= e^{-i\varepsilon_j t/\hbar}\Ket{\phi_j(t)}\,,
\end{align}
where $\varepsilon_j$ are known as quasienergies and $\Ket{\phi_j(t)}$ are the Floquet modes, and together are  the eigensystem of the Floquet operator ${\cal H}_{F}$,
\begin{align}
  \label{eq:HF_eigenvalue}{\cal H}_{F}\Ket{\phi_j(t)} &=\varepsilon_j\Ket{\phi_j(t)}\,,\\
  \label{eq:HF_operator}{\cal H}_{F} &= {\cal H}(t)-i\hbar\frac{\partial}{\partial t}.
\end{align}
Therefore, solving \eqref{eq:schrodinger_electronic} reduces to  find the Floquet modes. To accomplish this task, $\ket{\phi_j(t)}$ is expanded in a double Fourier series and in the basis of molecular states 
 which are eigenstates~$\ket{\zeta_l}$ of the electronic Hamiltonian ($\mathcal{H}_e\ket{\zeta_l}=V_l\ket{\zeta_l}$),
\begin{align}\label{eq:Floquet_mode}
    \Ket{\phi_j(t)} &= \sum\limits_{l}\sum\limits_{n_1,n_2=-\infty}^{\infty} C_{n_1,n_2}^{l,j}\ket{\zeta_l} e^{in_1\omega_1 t}e^{in_2\omega_2 t}\,.
\end{align}

Floquet modes, unlike molecular states, are not normalized for every time but only when they are averaged. For a single laser this is done over an optical cycle \cite{sambe_steady_1973}, but for two lasers with incommensurate frequencies a quasiperiodic average (see Appendix \ref{Appendix:Quasiperiodic_driving} for details) is needed, that is,
\begin{align}\label{eq:orthonormalization}
    \delta_{j^{\prime}j} &=\lim_{T\to\infty}\frac{1}{2T}\int\limits_{-T}^{T} \left\langle\phi_{j^{\prime}}(t)\vert \phi_j(t)\right\rangle dt = \timebraket[-0.05cm]{\!\!\phi_{j^{\prime}}}{\phi_j\!\!}.
\end{align}

The equation for the $C_{n_1,n_2}^{l,j}$ coefficients is derived in Appendix \ref{Appendix:Quasiperiodic_driving} (see Eq.~\eqref{eq:C_ljn1n2_ShirleyForm}) and is equivalent to the Hamiltonian in the Many-Mode Floquet Theory~\cite{ho1983}.

 At this point it is useful to introduce the extended space $H\otimes T_1\otimes T_2$ obtained by the direct product between the usual Hilbert space, the space of time-periodic functions for the first and second frequencies. This space is spanned by the direct product basis $\{ \Ket{\zeta_l,n_1,n_2}=\Ket{\zeta_l} \otimes \Ket{n_1} \otimes \Ket{n_2}\}$. The $C_{n_1,n_2}^{l,j}$ coefficients can be identified as the projection of the Floquet modes in this new basis so they can be written as
\begin{align}\label{eq:coefficients}
  C_{n_1,n_2}^{l,j}=\braket{\zeta_l,n_1,n_2|\phi_j}\,.
\end{align}

The original time-dependent system is now effectively a (time-independent) usual eigenvalue problem
\begin{align}\label{eq:H_F}
  \mathcal{H}_{F}
  \begin{pmatrix}
    \vdots\\
    c_{n_1,n_2}^{l,j}\\
    \vdots
  \end{pmatrix}
  &= \varepsilon_j
  \begin{pmatrix}
    \vdots\\
    c_{n_1,n_2}^{l,j}\\
    \vdots
  \end{pmatrix}\,,
  \end{align}
  where the diagonal elements of $\mathcal{H}_F$ are given by
  \begin{align}
  \label{eq:dressed_PES}\braket{\zeta_l,n_1,n_2|\mathcal{H}_F|\zeta_l,n_1,n_2} &= V_l+n_1\hbar\omega_1+n_2\hbar\omega_2\,,
  \end{align}
  and the off-diagonals
  \begin{align}
  \nonumber \braket{\zeta_\ell,n_1,n_2|\mathcal{H}_F|\zeta_l,m_1,m_2} &= \bra{\zeta_\ell}\big(E_1\delta_{n_1-m_1,-1}\delta_{n_2,m_2}\\
  \nonumber &+E_1^\dagger\delta_{n_1-m_1,1}\delta_{n_2,m_2}\\
  &+E_2\delta_{n_1,m_1}\delta_{n_2-m_2,-1}\\
  &+E_2^{\dagger}\delta_{n_1,m_1}\delta_{n_2-m_2,1}\big)\Ket{\zeta_{l}}\,.
\end{align}

Gu\'erin \textit{et al}. demonstrated the equivalence between Floquet formalism and the quantized version of the radiation field \cite{Guerin1997}, so $n\hbar\omega_1$ and $n\hbar\omega_2$ in \eqref{eq:dressed_PES} can be interpreted as the photon energy that is emitted ($n>0$) or absorbed ($n<0$) from the field (see also section II B in \cite{shirley1965} and part II chapter 4 in \cite{joachain_2011}).

Here we use the \ce{Cs_2} molecule as a model system in our calculations. For the sake of simplicity, we focus only on the low-lying $\Sigma$ states of \ce{Cs_2}, i.e, $X^1\Sigma_g^{+}$, $A^1\Sigma_u^{+}$ and $B^1\Sigma_g^{+}$, that we will denote as $\Ket{\zeta_a}$, $\Ket{\zeta_b}$ and $\Ket{\zeta_c}$ respectively. These states form a ladder scheme transition, owing to the absence of transition dipole moment (TDM) between $X^1\Sigma_g^{+}$ and $B^1\Sigma_g^{+}$ (transition forbidden by symmetry). Coupling $\Ket{\zeta_a}$ with $\Ket{\zeta_b}$ and $\Ket{\zeta_b}$ with $\Ket{\zeta_c}$ through the absorption of one photon $\hbar\omega_1$ and $\hbar\omega_2$ respectively, the Floquet Hamiltonian reads
\begin{align}\label{eq:H_3x3}
  \mathcal{H}_{F} &=
  \begin{pmatrix}
  V_a(R) & H_{ab}(R) & 0 \\
  H_{ab}^{\ast}(R) & V_b(R)-\hbar\omega_1 & H_{bc}(R) \\
  0 & H_{bc}^\ast(R) & V_c(R)-\hbar\omega_1-\hbar\omega_2
  \end{pmatrix},\\
  \label{eq:off-diagonal}H_{xy}(R) &= \frac{1}{2}\epsilon_0^{(k)} d_{xy}(R)\sin{(\theta_k)}e^{i\delta_k}\,,\quad
  \begin{cases}
  \quad k=1,2\\
  x,y=\{a,b,c\}
  \end{cases}\,,
\end{align}
where $d_{xy}(R)$ is the transition dipole moment between molecular states $\Ket{\zeta_x}$ and $\Ket{\zeta_y}$; $\theta_k$ is the angle between the molecular axis and the propagation direction of the electric field, and $\delta_k$ in the case of linearly (circularly) polarized light is the initial phase (orientation of the molecule in the polarization plane) defined in \eqref{eq:polarization}. Equation~\eqref{eq:H_3x3} implies that both, the distance between atoms and the orientation of the molecule are taken as parameters. If the rotation of the molecule with respect to the laser fields were to be taken as a dynamical variable, then transition dipole moments in the molecular frame should be rotated to the laboratory frame, which is usually the frame in which the polarization of light is defined. For instance, while in the molecular frame $d_{xy}=\bra{\zeta_x}r\Ket{\zeta_y}$, in the laboratory frame $d^{lm}_{xy}=\bra{\zeta_x}r^{\prime}Y_{1m}(\theta^{\prime},\phi^{\prime})\Ket{\zeta_y}$, where $Y_{1m}$ are spherical harmonics in the in the laboratory frame \cite{Cherepkov1981,demekhin2013}. For all our calculations, the  BO potential energy surfaces and transition dipole moment in the molecular frame were taken from \cite{allouche2012}.

\section{Results and Discussion}\label{sec:resonances}
\subsection{Resonances}
Degeneracies in molecular systems are ubiquitous, with conical intersections (CIs) between two states being the most well-known \cite{yarkony2001}. However, for diatomic molecules the non-crossing rule does not allow a degeneracy between the PESs to occur \cite{Naqvi1972}, because only one degree of freedom (DoF) is available (interatomic distance $R$) and at least two are required in order to obtain a CI. Instead, avoided crossings (ACs) emerge, such as the one seen between the $1^1\Sigma^{+}$ and $2^1\Sigma^{+}$ states of LiF \cite{werner1981}. 
CIs and ACs are not restricted to the Born-Oppenheimer (BO) picture but can also be induced by external fields \cite{wallis2009,arasaki2010} and take place even in diatomic molecules \cite{garraway1998,moiseyev2008}, forming the so-called light-induced conical intersections (LICIs) \cite{Sindelka2011}. The conditions for a LICI to emerge are simple, the coupling between electronic states must vanish and the laser has to be resonant with the energy differences between states~\cite{demekhin2013}. In our case, in which the orientation is taken as a parameter, there will be a LICI if $V_a(R)=V_b(R)-\hbar\omega_1$, $V_b(R)=V_c(R)-\hbar\omega_2$ and the laser propagates parallel to the TDM~\cite{Gershnabel2006,chang2019,Carrasco2022}. This case is somewhat trivial because in the molecular frame a field propagating parallel to the TDM does not couple with matter. This, however, is a consequence of not treating rotation as a dynamical variable. When this is done, one finds that the effective coupling between electronic states cancels out for less trivial orientations. A detailed discussion of the conditions of LICIs formations can be found in Demekhin's work~\cite{demekhin2013}. Note also that in our treatment the formation of a triple LICI does not depend on the polarization of light as the term $\sin{(\theta_k)}e^{i\delta_k}$ in the Hamiltonian \eqref{eq:H_3x3} could be taken as an effective orientation.

On the other hand, once this effective orientation is different from $0$ or $\pi$, the LICI is lifted and replaced by a light-induced avoided crossing (LIAC). The way energy gaps opens depends on the strength of the fields and the states involved in the crossing, but the general picture is that the locations of the LIACs are controlled mostly by the frequency of the laser while the gaps are controlled by the intensity of the fields. 

\begin{figure}[ht!]
  \centering\includegraphics[width=\columnwidth]{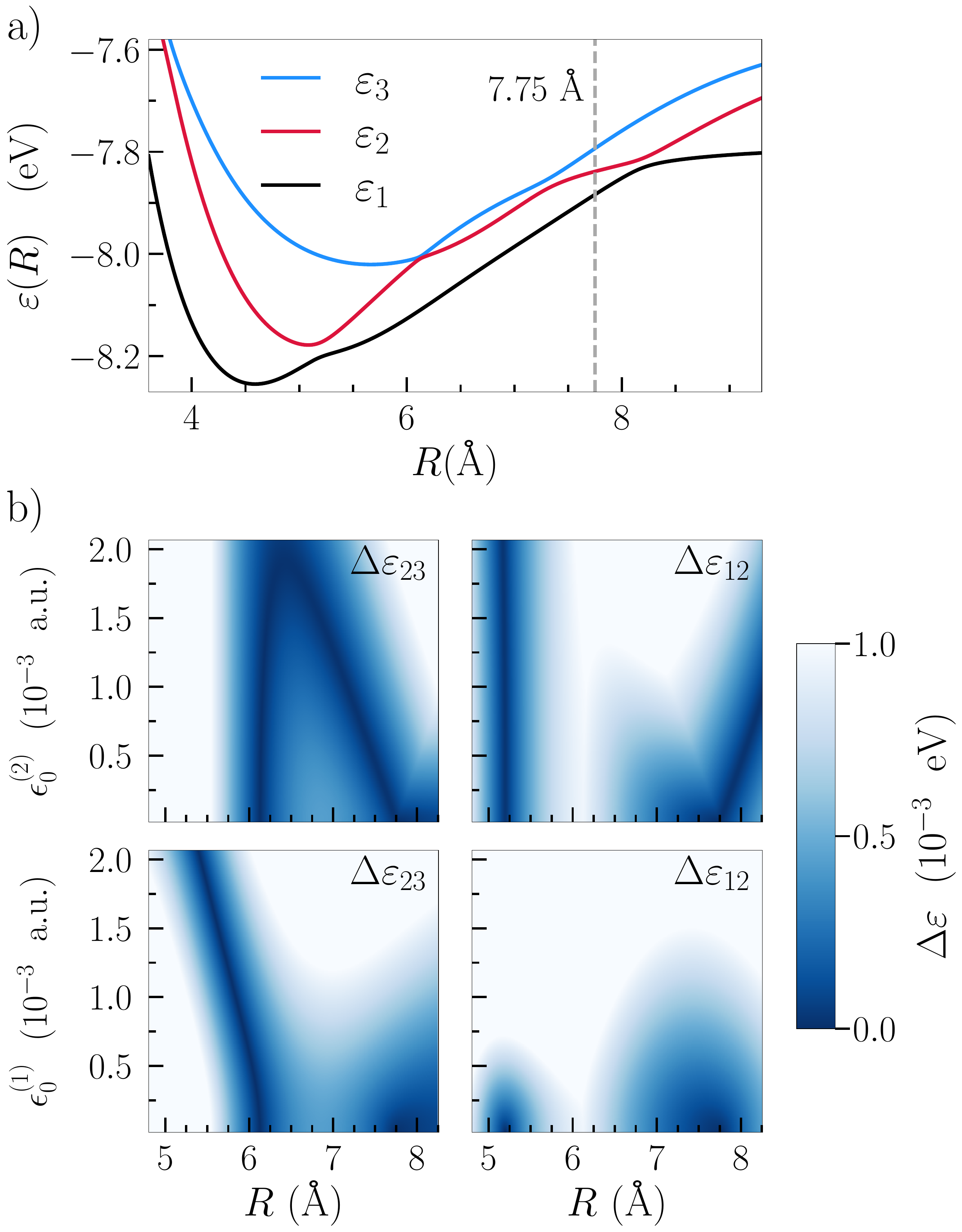}
  \caption{a) Quasienergies of $Cs_2$ interacting with two laser of field strengths  $\epsilon_0^{(1)}=0.0002$ and $\epsilon_0^{(2)}=0.0008$ a.u. b) Dependence of quasienergies gap with the field strength. Top row: gap of the avoided crossing between states 2 and 3 and 1 and 2 as function of $\epsilon_0^{(1)}$ while $\epsilon_0^{(2)}$ is fixed to an small value of $0.000017$ a.u. ($I=10^{7}$ W/cm$^2$). Bottom row: same as in top row but now $\epsilon_0^{(1)}$ is varied while $\epsilon_0^{(2)}$ is kept fixed to $0.000017$ a.u.}
  \label{fig:splitting}
\end{figure}

If both lasers are simultaneously resonant at some interatomic distance $R_0$, a three-level crossing appears between the states $\Ket{\zeta_a,0,0}$, $\Ket{\zeta_b,-1,0}$ and $\Ket{\zeta_c,-1,-1}$. In this scenario, we aim to investigate the behavior of the LIACs of $\mathcal{H}_F$ \eqref{eq:H_3x3} as we vary the fields strengths. As a concise example, we chose frequencies $\hbar\omega_1= 1.12398$ eV and $\hbar\omega_2= 0.16054$ eV, which create a three-level crossing at $R=7.75$ \AA\ (see Figure \ref{fig:splitting} a)). Other crossings will also be present at different interatomic distances but involving two states. This happens between $\Ket{\zeta_a,0,0}$ and $\Ket{\zeta_c,-1,-1}$ at $R=6.13$ \AA, and with $\Ket{\zeta_b,-1,0}$ at $R=5.2$ \AA. The resonance between $\Ket{\zeta_b,-1,0}$ and $\Ket{\zeta_c,-1,-1}$ occurs in the repulsive region ($R=3.74$ \AA) where the Pauli repulsion dominates.

\begin{figure}[ht!]
  \centering
  \includegraphics[width=\columnwidth]{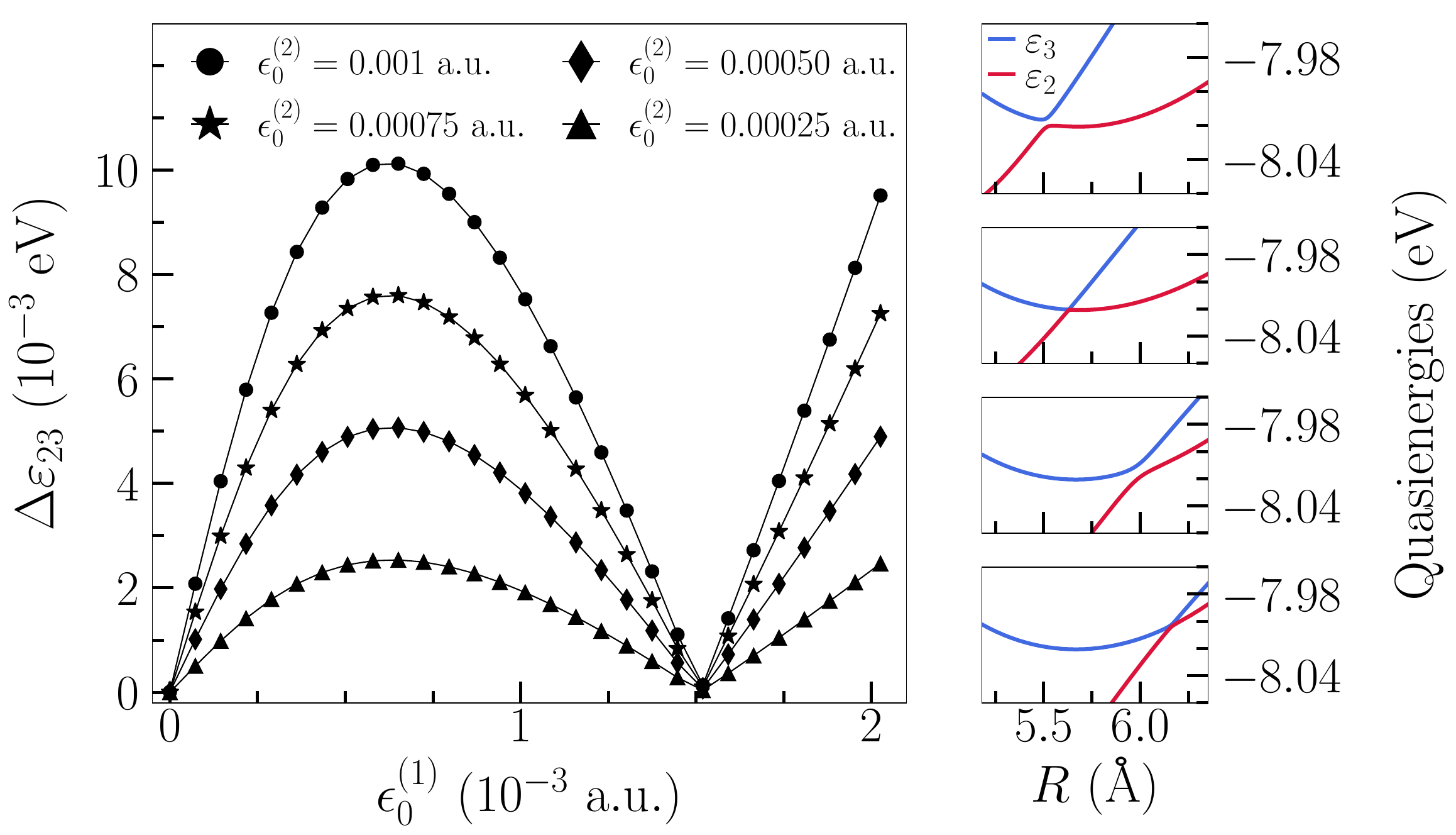}
  \caption{Gap between $\varepsilon_2$ and $\varepsilon_3$ as function of field strength. The left panels show the quasienergies for $\epsilon_0^{(2)}=1.0\times10^{-3}$ a.u. and $\epsilon_0^{(1)}=0.5\times10^{-4}$, $\epsilon_0^{(1)}=0.6\times10^{-3}$, $\epsilon_0^{(1)}=1.5\times10^{-3}$, $\epsilon_0^{(1)}=1.8\times10^{-3}$ a.u. in ascending order respectively.} 
   \label{fig:gap}
\end{figure}

To explore the response of the LIACs as we vary $\epsilon_0^{(k)}\sin(\theta_k)$ ($k=1,2$), we studied the quasienergy gaps $\Delta\varepsilon_{nm}$. For clarity, hereafter the orientation will always be implicitly included in this product $\epsilon_0^{(k)}\sin(\theta_k)$, which we will simply call the field strength $\epsilon_0^{(k)}$. A counter-intuitive aspect stands out in the energy gaps of the two LIACs originated from the three-level crossing, they are not lifted vertically but drift to the right and left from the position of the LICI, $R=7.75$ \AA\ (see Figure \ref{fig:splitting} a)). This peculiarity is not exclusive of this crossing. The LIAC associated with the intersection of $\Ket{\zeta_a,0,0}$ and $\Ket{\zeta_c,-1,-1}$ also shows this behaviour. In the upper (lower) left panel of Figure \ref{fig:splitting} b) we can observe that this LIAC moves to the right from its original location as we increase $\epsilon_0^{(2)}$ with $\epsilon_0^{(1)}$ fixed. The opposite occurs when $\epsilon_0^{(1)}$ is varied while $\epsilon_0^{(2)}$ is kept constant, as it can be seen in the lower left panel of Figure \ref{fig:splitting} b). Interestingly, as the LIAC between states $\ket{\phi_2}$ and $\ket{\phi_3}$ moves to shorter interatomic distances, a critical value of $\epsilon_0^{(1)}$ is always achieved at which the gap closes independently of the value of $\epsilon_0^{(2)}$. This feature takes place when the former LIAC is moving to the left (see Figure \ref{fig:gap}). This degeneracy always occurs at $R= 5.6$ \AA\ and it is accidental in the sense that it arises because at this distance the TDM between $\Ket{\zeta_b}$ and $\Ket{\zeta_c}$ is zero causing  $H_{bc}$ in Eq.\eqref{eq:off-diagonal} to vanish.
Another interesting feature in Figure \ref{fig:gap} is the quadratic nature of the gap between $\Ket{\zeta_a}$ and $\Ket{\zeta_c}$. This is because, although those states have a null TDM, $\Ket{\zeta_b,-1,0}$ acts as an intermediary state.

\subsection{Initial phase and transition probabilities}\label{sec:Initla_Phase}

Hitherto we have explored two out of three parameters that define the driving force: frequencies and fields strengths. The remaining parameter is the initial phase of the lasers $\delta_j$, and in this section we show that the transition probabilities between molecular states could depend on $\delta_j$. In Floquet Theory, the passage of time operates on two distinct scales. The first scale, known as stroboscopic evolution, is governed by quasienergies ($\hbar/\varepsilon$). In the case of a single driving force, this scale involves time intervals that differ by an integer number of periods, giving it its name. The second scale, micromotion, refers to much shorter time intervals compared to the driving period. The stroboscopic transition probabilities~\cite{ho1983},
\begin{align}
\label{eq:stro-prob}
  P_{f\gets i}^{(S)} &= \sum_{n_1,n_2}\left|\braket{\zeta_f,n_1,n_2|e^{-i\mathcal{H}_Ft/\hbar}|\zeta_i,0,0}\right|^2\,,
\end{align}
would no depend on the phase when the frequencies are incommensurate because in that case the quasienergies do not depend on it either \cite{dorr1991,chu2004}. This is not necessarily true for commensurate frequencies with small  $\omega_1 /\omega_2$ ratio, as it have been discussed by Poertner \textit{et al.}~\cite{poertner2020} and Sirko \textit{et al.}~\cite{sirko2001}. If we stick to incommensurate ratios $\omega_1/\omega_2$ , as the one used in previous sections, the effects of the initial phase, if any, will only manifest through the micromotion. 

If the system is initially in the molecular ground state $\Ket{\zeta_a}$, the transition probability to other molecular state $\Ket{\zeta_f}$ at an arbitrary time $t$, that is including the micromotion, is given by~\cite{ho1984}:

\begin{align}\label{eq:prob_amplitude}
  P_{f\gets a} &= \left |\sum_{n_1,n_2} e^{i(n_1\omega_1+n_2\omega_2)t}\braket{\zeta_f,n_1,n_2|e^{-i\mathcal{H}_F t/\hbar}|\zeta_a,0,0} \right |^2\,.
\end{align}

The dependency of $P_{f\gets a}$ on the initial phase is implicitly encoded in the Floquet Hamiltonian. Although this probabilities will in general depend on the phase, a minimal basis set is not enough to capture this dependency and replicas that allow the absorption and emission of multiple photons are needed. For instance, in the minimal-basis $\mathcal{H}_F$ \eqref{eq:H_3x3}, the phase can be eliminated with an unitary transformation,
\begin{align}
  U &=
  \begin{pmatrix}
    e^{-\delta_1} & 0 & 0\\
    0 & e^{-2i\delta_1} & 0 \\
    0 & 0 & e^{-i(2\delta_1+\delta_2)}
  \end{pmatrix}\,.
\end{align}

\begin{figure}[ht!]
  \centering
  \includegraphics[width=\columnwidth]{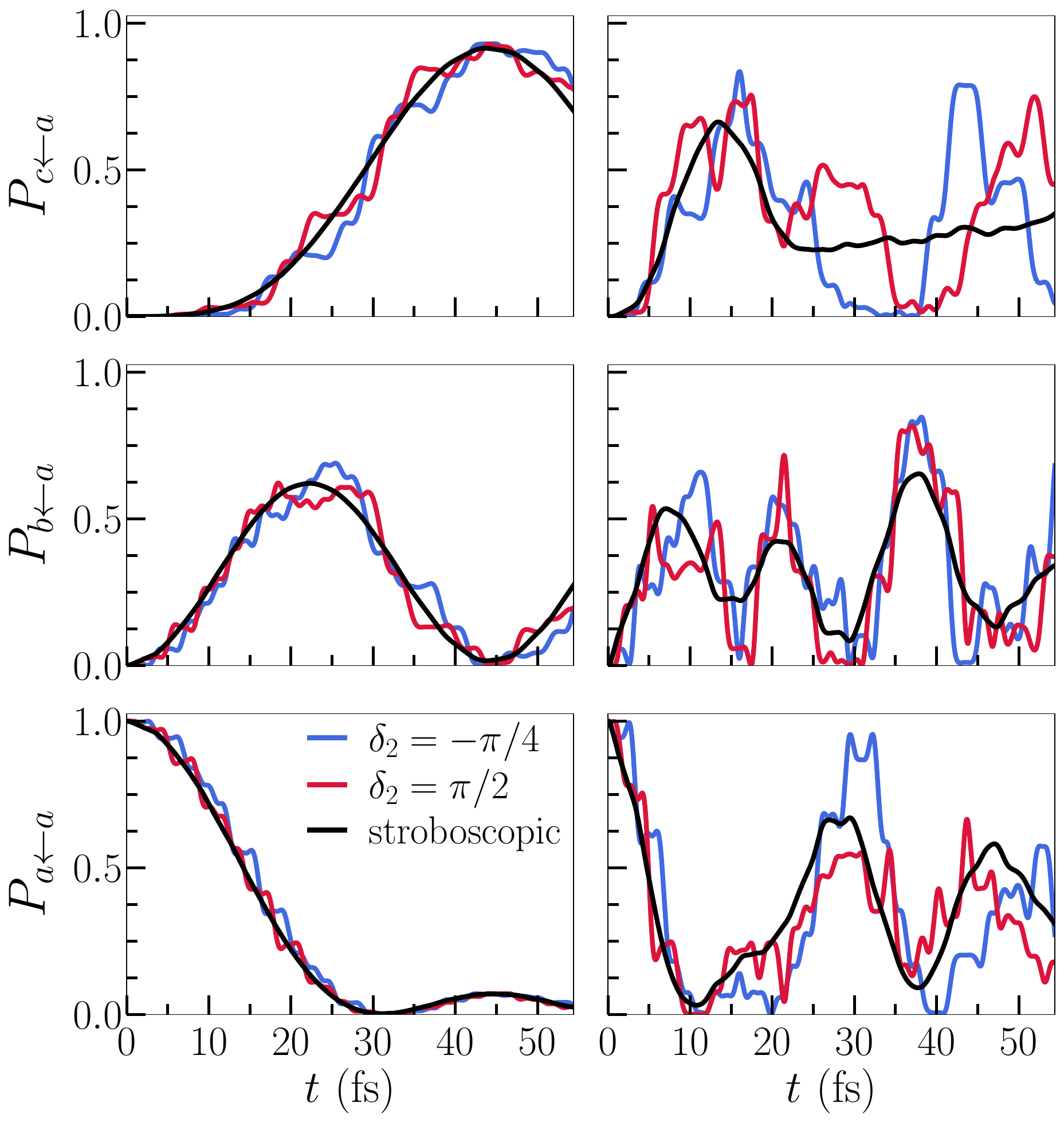}
  \caption{Transitions probabilities between molecular states as function of time calculated at $R= 7.75$ \AA\ with the phase of one laser fixed at $\delta_1=\pi/4$  and two values of the phase of the second  laser:  $\delta_2=-\pi/4$  in blue and  $\delta_2=\pi/2$ in red.  Probabilities include the micromotion  in the time-evolution operator (Eq. \eqref{eq:stro-prob}). The black line corresponds to the stroboscopic evolution only (Eq. \eqref{eq:prob_amplitude}). Left column:  intensity $I_1=I_2=I=10^{10}$ W/cm$^2$ . Right column:  intensity $I_1=I_2=I=10^{11}$ W/cm$^2$ .}
  \label{fig:transitions}
\end{figure}

Therefore, we extended the basis of the Hamiltonian of Eq.\eqref{eq:H_3x3} to include harmonics $n_1=0,\ \pm1,\ldots, \pm4$ and $n_2=n_1\pm 2n\,$, $n=1,\ldots,4\,$ ($\ket{\zeta_a,n_1,n_2}$, $\ket{\zeta_b,n_1-1,n_2}$ and $\ket{\zeta_c,n_1-1,n_2-1}$). This results in a $243\times243$ Hamiltonian which ensures norm-conservation of the wavefunction within $10^{-3}$.  Figure \ref{fig:transitions} shows the transition probabilities $P_{c\gets a}$, $P_{b\gets a}$ and population $P_{a\gets a}$ for two phases and two laser intensities. The stroboscopic probability is also show with a solid-black curve. Numerical tests show that the instantaneous probabilities (Eq. \eqref{eq:prob_amplitude}) start to deviate from the stroboscopic ones (Eq.\eqref{eq:stro-prob}) at intensities of about $I=10^{10}$ W/cm$^2$. At this intensity the coupling strengths ($g_j=|H_{xy}|^2/(\hbar\omega_j)^2$, $xy =ab,\ bc$) are $g_1=1.25\times 10^{-3}$ and $g_2=3.26\times 10^{-2}$, could be considered in the borderline between moderate and strong coupling. Moreover, if one were interested in an experiment in which the lasers remain on for long times and the coupling is weak or moderate, the time-averaged transition probabilities would not be too different from the average of the stroboscopic probabilities~\cite{ho1983} 
\begin{align}\label{eq:average_P}
  \overline{P}_{f\gets a} &= \sum_{j,n_1,n_2}\left|\braket{\zeta_f,n_1,n_2|\varepsilon_j}\braket{\varepsilon_j|\zeta_a,0,0}\right|^2\,.
\end{align}

In contrast, when the intensity reaches  $I=10^{11}$ W/cm$^2$, the system enters the strong coupling regime and the coupling parameters are  $g_1=1.25\times 10^{-2}$ and $g_2=3.26\times 10^{-1}$. In this regime, the transition probabilities are significantly influenced by the initial phase and deviate significantly from stroboscopic evolution. In some cases, transitions can be hindered (observed in $P_{a\gets c}$   between 30 and 40  fs) or amplified (observed in $P_{a\gets a}$  between 25 and 35 fs).  The observed population transfer  points up that defining a potential energy surface  at the \textit{micromotion} time scale is not meaningful. Additionally, these population changes happen within a short time-scale (approximately 10 fs), which is considerably faster than the lowest vibrational period of the ground state of Cs$_2$ (740 fs, as reported by Allouche in 2012 \cite{allouche2012}) and, in fact, faster than the period of most molecules. Consequently, the initial phase can effectively introduce non-adiabatic behavior into nuclear dynamics within this fast timescale.

\subsection{Light-Induced Potentials}

So far we have suggested that the quasienergies $\varepsilon_j$ can be interpreted as potential energy surfaces, i.e., they represent a mean field for nuclear dynamics. In fact, it is common in the literature to refer to $\varepsilon_j$ as light-induced potentials (LIPs) \cite{corrales2014,coffee2006,hanasaki2013,kubel2020}. A proof that $\varepsilon_j$ plays the role of a mean field is necessary, as well as a discussion of the limits of this interpretation. We start by writing the total nuclear and electronic wavefunction as a linear expansion in the basis of Floquet modes
\begin{align}\label{eq:nuclear_floquet_ansatz}
  \Ket{\Psi (R,r,t)} &=\sum_{j}\chi_j(R,t) \Ket{\phi_j(r,t)}\,.
\end{align}
In Equation \eqref{eq:nuclear_floquet_ansatz} the interaction of the electrons with the field is captured by Floquet modes, while the nuclear dynamics is encoded in the expansion coefficients $\chi_j(R,t)$. Note the similarity between equation \eqref{eq:nuclear_floquet_ansatz} and the Born-Huang representation of the time-dependent molecular wavefunction in the absence of a radiation field
\begin{align}\label{eq:Born-Huang_expansion}
  \Ket{\Psi (R,r,t)} &=\sum_{j}\chi^{\rm (BH)}_j(R,t) \Ket{\zeta_j(r)}\,.
\end{align}

The only difference between Eq.~\eqref{eq:nuclear_floquet_ansatz} and the Born-Huang representation \eqref{eq:Born-Huang_expansion}, is that in the latter the BO molecular states do not depend on time. In both cases, the expansion coefficients $\chi_j(R,t)$ are nuclear wavefunctions. To find the time-evolution of $\chi_j(R,t)$ one needs to replace Eq.~\eqref{eq:nuclear_floquet_ansatz} in the time-dependent Schr\"odinger equation \eqref{eq:schrodinger}. Using  Eq.~\eqref{eq:HF_eigenvalue} and the completeness of the time-averaged Floquet modes (see Appendix \ref{Appendix:derivative_couplings} for a detailed derivation), one finds an Schr\"odinger equation for the nuclear wavefunction

\begin{align}\label{eq:nuclearTDSE}
    \nonumber i\hbar\frac{\partial \chi_k}{\partial t} &=-\frac{\hbar^2}{2\mu} \nabla^2\chi_k+\varepsilon_k\chi_k-\frac{\hbar^2}{2\mu}\sum\limits_{j}\chi_j\timebraket[-0.08cm]{\!\!\phi_j\!}{\!\nabla^2\phi_k\!\!}\\
     &-\frac{\hbar^2}{\mu}\sum\limits_{j}\nabla\chi_j\cdot\timebraket[-0.05cm]{\!\!\:\phi_j\!}{\!\nabla\phi_k\!\!}\,,
\end{align}
where $\mu$ stand for the reduced mass, and $\timebraket[-0.05cm]{\!\!\:\phi_j\!}{\!\nabla\phi_k\!\!}$ and $\timebraket[-0.08cm]{\!\!\phi_j\!}{\!\nabla^2\phi_k\!\!}$ are the first and second order non-adiabatic derivative couplings, respectively. 

Equation \eqref{eq:nuclearTDSE} is formally equivalent to the nuclear Schr\"odinger equation in the Born-Huang picture, though with some distinctions. In this formulation, the couplings elements account for both light-induced effects and those inherent in the field-free molecular states. 
It is important to note that Equation \eqref{eq:nuclearTDSE} applies only in an averaged timeframe. Specifically, it accurately describes nuclear dynamics over longer periods, reflecting the system's stroboscopic evolution. Shorter timescales, such as micromotions occurring within a single period, are not captured by Equation \eqref{eq:nuclearTDSE}. Consequently, the quasienergies $\varepsilon_j$ serve as mean-fields for nuclear motion (depicting potential energy surfaces) primarily in regions where non-adiabatic effects are insignificant and when the scale of nuclear movement greatly surpasses the driving period, a criterion commonly met. For example, Cs$_2$ in its ground state exhibits an oscillation period of $749$ fs, while the period of ultraviolet radiation is shorter than $1$~fs.

To separate the light-induced nonadiabaticities from those
of the molecular states, it is convenient to expand the
nonadiabatic derivative couplings in the basis of the extended space (Eq.~\eqref{eq:Floquet_mode})

\begin{align}
    \nonumber\Tau_{kj} &= \timebraket[-0.05cm]{\phi_k\!}{\nabla\phi_j\!}\\
    \nonumber &=\sum\limits_{l,n_1,n_2} C_{n_1,n_2}^{l,k\!\:\ast}\nabla C_{n_1,n_2}^{l,j}\\
    \label{eq:first_coupling} &+\sum\limits_{\substack{l,\ell\\n_1,n_2}}C_{n_1,n_2}^{l,k\!\:\ast} C_{n_1,n_2}^{l,j}\Braket{\zeta_l|\nabla\zeta_\ell}\,,\\
    \nonumber\Tau_{kj}^{(2)} &= \timebraket[-0.08cm]{\phi_k\!}{\nabla^2\phi_j\!}\\
    \nonumber &=\sum\limits_{l,n_1,n_2} C_{n_1,n_2}^{l,k\!\:\ast}\nabla^2 C_{n_1,n_2}^{l,j}\\
    \nonumber &+\sum\limits_{\substack{l,\ell\\n_1,n_2}}C_{n_1,n_2}^{l,k\!\:\ast} C_{n_1,n_2}^{\ell,j}\Braket{\zeta_l|\nabla^2\zeta_\ell}\\
    \label{eq:second_coupling}&+2\sum\limits_{\substack{l,\ell\\n_1,n_2}}C_{n_1,n_2}^{l,k\!\:\ast} \Braket{\zeta_l|\nabla\zeta_\ell}\cdot\nabla C_{n_1,n_2}^{\ell,j}
\end{align}

In the first \eqref{eq:first_coupling} and second \eqref{eq:second_coupling} order derivative couplings one can identify those induced by light as those terms in which no derivatives of the molecular states with respect to the nuclear positions appear,
\begin{align}
    \tau_{kj}^{\rm LI } &= \sum\limits_{l,n_1,n_2} C_{n_1,n_2}^{l,k\!\:\ast}\nabla C_{n_1,n_2}^{l,k\!\:\ast}\,,\\
    \tau_{kj}^{{\rm LI} \!\:(2) } &= \sum\limits_{l,n_1,n_2} C_{n_1,n_2}^{l,k\!\:\ast}\nabla^2 C_{n_1,n_2}^{l,k\!\:\ast}\,.
\end{align}

On the other hand, since light couples all molecular states, the non-adiabatic coupling between Floquet modes $k$ and $j$ involves all the couplings between field-free sates, $\Braket{\zeta_l|\nabla\zeta_\ell}$, $\Braket{\zeta_l|\nabla^2\zeta_\ell}$,
\begin{align}
    \tau_{kj}^{\rm Mol } &= \sum\limits_{\substack{l,\!\:\ell\\ n_1,n_2}} C_{n_1,n_2}^{l,k\!\:\ast}\Braket{\zeta_l|\nabla\zeta_\ell} C_{n_1,n_2}^{\ell,j}\,,\\
    \tau_{kj}^{{\rm Mol} \!\:(2) } &= \sum\limits_{\substack{l,\!\:\ell\\ n_1,n_2}} C_{n_1,n_2}^{l,k\!\:\ast}\Braket{\zeta_l|\nabla^2\zeta_\ell} C_{n_1,n_2}^{\ell,j}\,,\\
    \tau_{kj}^{\rm LI-Mol } &= \sum\limits_{\substack{l,\ell\\n_1,n_2}}C_{n_1,n_2}^{l,k\!\:\ast} \Braket{\zeta_l|\nabla\zeta_\ell}\cdot\nabla C_{n_1,n_2}^{\ell,j}\,.
\end{align}

Finally, it is important to mention that the ansatz of Equation \eqref{eq:nuclear_floquet_ansatz} is not the only one possible. For example, Schir\`o  \textit{et al.}~\cite{schiro2021} recently developed a trajectory-based approximation to describe the molecular dynamics of driven systems using the Floquet formalism and the exact factorization~\cite{abedi2010} of the total wave function, while Sch\"afer \textit{et al}. also used the Born-Huang ansatz but kept the radiation field quantized\cite{schafer2018}.

\subsection{Photoassociation}

Cold and ultracold molecules refer to those with very low translational energy, typically in the range of nanokelvin to millikelvin~\cite{weiner1999,friedrich2009}. At such temperatures, a molecule's de Broglie wavelength can be significantly larger than its actual size. Consequently, molecules in this state can exist in various quantum states, spanning from vibrational to hyperfine levels. Cold molecules offer a unique opportunity to study atomic and molecular interactions, as well as chemical reactivity, in great detail~\cite{balakrishnan2016,zhang2023}. There are two main methods for creating cold molecules: direct and indirect. Direct methods involve extracting the translational kinetic energy from hot molecules, whereas indirect methods resort to the association of previously cooled atoms. Indirect methods are generally simpler because the intricate ro-vibrational states of molecules make direct cooling with lasers challenging. One specific approach, called the photoassociation method, involves creating a molecule by colliding two ultracold atoms in the presence of resonant lasers~\cite{fioretti2000}.

\begin{figure}[ht!]
  \centering
  \includegraphics[width=\columnwidth]{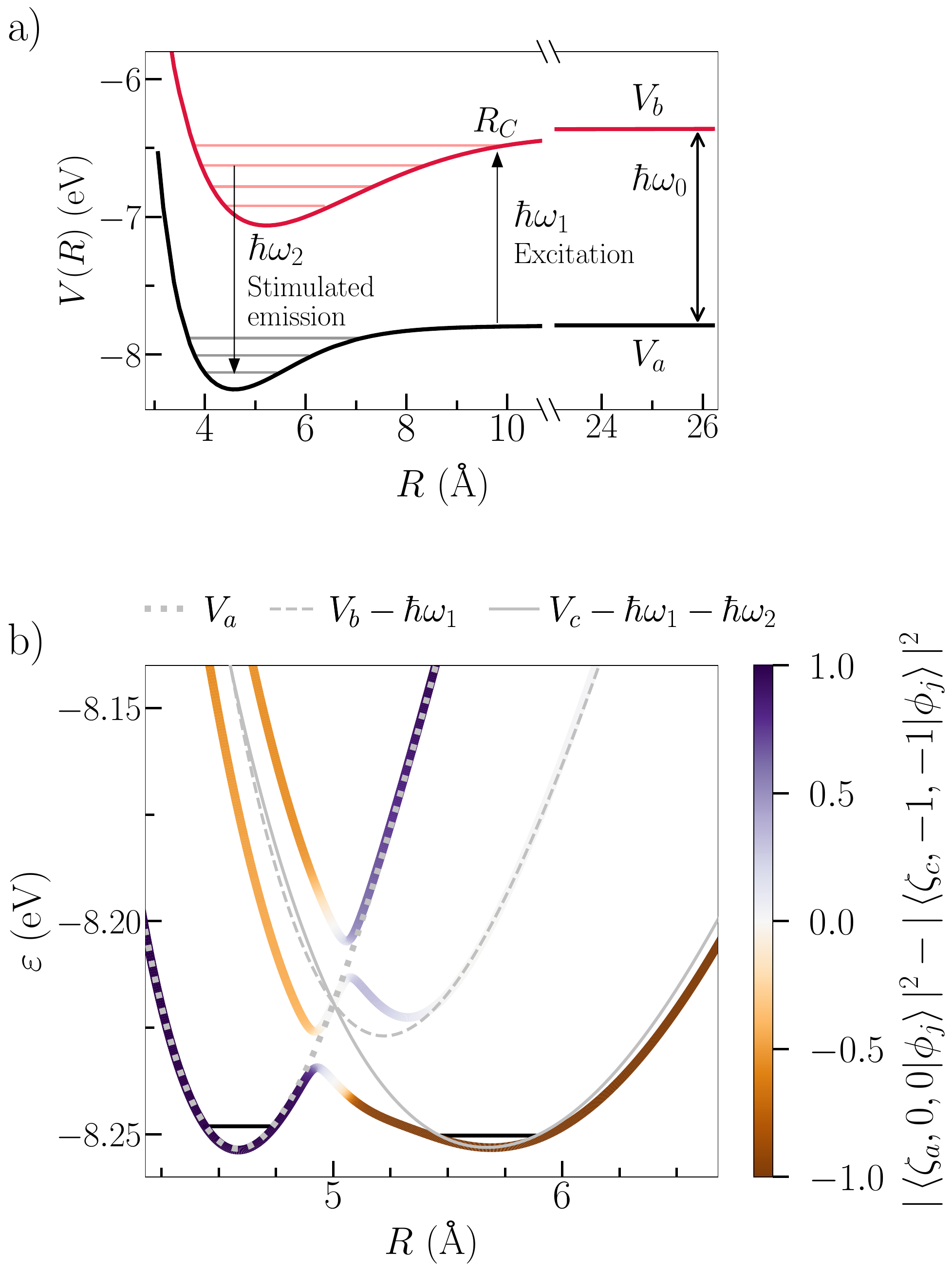}
  \caption{a) Sketch of a usual photoassociation scheme in which a laser of  energy $\hbar\omega_1$  brings the atoms to a vibrationally excited Condon state at $R_C$  while a second laser brings the molecule to the cold ground state by stimulated emission \cite{friedrich2009} . b) An alternative route for  photoassociation based on the engineering   of potential energy surfaces   of $Cs_2$ (see text) .    Two laser with $\hbar\omega_1=1.164$ eV, $\hbar\omega_2=0.353$ eV, $I_1=2.5\times10^{8}$ and $I_2=2.5\times10^{10}$ W/cm$^2$ create a ground state with  a double-well shape in which a molecule trapped in the long-range well can tunnel  to the short-range well of the molecular ground state.  An orange color means that the largest composition of the Floquet mode is the second molecular excited state while blue that the largest contribution is from the ground molecular state.  
  Horizontal  black lines at the bottom of the wells represent the lowest vibration level of $V_a$ ($X^1\Sigma_g^{+}$) and $V_c$ ($B^1\Sigma_g^{+}$), $45$ cm$^{-1}$ and $22.62$ cm$^{-1}$ respectively \cite{allouche2012}.}
  \label{fig:double_well}
\end{figure}

In the ground state, two cold atoms cannot form a molecule due to the absence of a thermal mechanism that releases the binding energy without increase the translational energy. Moreover, the potential energy in the dissociative regime of the ground state follows a London $1/R^6$ trend. Consequently, when cold atoms meet at large distances, there are no significant forces to facilitate their collision. However, in excited states, the potential energy of bound states decreases according to a $1/R^3$ pattern, forming bonding wells at greater distances (see Figure~\ref{fig:double_well} a)). This results in attractive forces operating over larger distances. Photoassociation capitalizes on this behavior of the excited states' potential energy surface \cite{weiner1999}. Initially, the system is excited using a laser with energy $\hbar\omega_1$, which is red-detuned relative to the atomic excitation energy $\hbar\omega_0$. This excitation leads to a free-bound transition around the Condon point $R_C$, resulting in the formation of electronically and vibrationally excited molecules. If collision occurs between these vibrationally excited molecules, kinetic energy transfers to the translational degree of freedom, potentially leading to the formation of warm molecules. However, an additional laser, operating at the appropriate frequency, $\hbar\omega_2$, could induce stimulated emission, causing the molecules to return to the electronic and vibrational ground state. Remarkably, throughout this process, the formed molecule retains the same translational energy as the parent cold atoms \cite{jones2006}.

In our case, in which the molecule is always illuminated with two lasers, we have the necessary ingredients for photoassociation: \textit{i)} the ability to modify the PES in order to create long-range bonding wells, and \textit{ii)} the possibility of having some degree of control over the population transfer between molecular states (section \ref{sec:Initla_Phase}). For instance, in Figure~\ref{fig:double_well} b) we crafted PES's with a convenient double-well shape by using a laser red-detuned ($0.2628$ eV) to the common dissociation limit of both excited states ($6s ^2S+ 2p ^2P$) plus a second laser with $\hbar\omega_2=0.353$ eV. The molecule-radiation ground state has a long-range bonding well with a minimum around $6$ \AA\ which provides an association channel for distant atoms. The shape of this well is dominated by the field-free PES of the second excited state ($B^1\Sigma_{g}^+$) . Note that if only one laser were used, this state would not be accessible from the ground state as this transition is symmetry forbidden. This is one of the advantages of using bichromatic radiation since the second laser mediates the transition to more excited states, which, as in our case, may have long-range bonding wells. However, if the lasers were turned off while the atoms vibrate around 6 \AA, the molecule would decay to the molecular ground state ($A^1\Sigma_{g}^+$) in a Franck-Condon region corresponding to vibrational excited state (warm molecule). Nonetheless, the double-well  shape of the ground state effective PES, with a small barrier ($0.0187$ eV) and almost-resonant vibrational ground states of the left and right wells favors the tunnelling between them. With this, a significant fraction of molecules would populate the left bonding well, which is essentially the well of the molecular ground state (blue col). Another advantage of the two-lasers setup is that the population of the molecular ground state can be controlled with the phase difference or delay of the lasers. We would like to call the attention to the theoretical work by Pawlak \textit{et al}.~\cite{pawlak2015}, in which the authors investigated the binding of cold Rb atoms in the periodic potential of a 1-D optical lattice, showing that the non-adiabatic effects near a LICI enhances the localization (association) of the wavefunctions of Rb$_2$ molecules.

\section{Conclusions and final remarks}

In this study, we explored the interactions between two lasers of differing frequencies and a Cs$_2$  molecule, employing many-mode Floquet theory to probe the intricacies of the ground and excited states of the molecule. Our research has led to several insights into the quantum dynamics of molecular systems under electromagnetic fields.

Each laser interaction effectively introduces an additional dimension to the Hamiltonian of the Cs$_2$ molecule. This addition allows us to lift the limitations  associated with the non-crossing rule for adiabatic molecular states, resulting in the formation of Light-Induced Conical Intersections (LICI) and a variety of avoided crossings among hybridized radiation-matter states. The rotation of the molecule, relative to the light polarization, was treated as a static parameter, revealing that LICIs are formed only when the field aligns with the molecular transition dipole moment.

Additionally, our findings elucidate the significant influence of field strength and orientation on the characteristics of avoided crossings. These factors not only determine the size of the energy gap but also influence the position of these gaps along the interatomic distance, introducing a dynamic element to the resonance conditions and the possibility of accidental degeneracies at points where the transition dipole moment is zero.

The role of the initial phase difference between the two lasers on the transition probabilities between molecular states was also examined. At a larger timescale, known as stroboscopic motion, the phase difference does not impact these probabilities. However, at the scale of micromotion, the initial phase can modulate transition probabilities, suggesting a method to manipulate non-adiabatic dynamics that occur over timescales shorter than nuclear motion.

We have provided a detailed discussion on the interpretation of quasienergies in relation to nuclear positions. These can be understood as light-induced potential energy surfaces for the nuclear motion  in the presence of light, but such an interpretation holds true only when averaging over the fast oscillations of  the micromotion. Furthermore, the interaction with the radiation fields introduces various non-adiabatic couplings between these quasienergy surfaces, encompassing inherent system couplings, those induced by the light, and a combination of the two.

Finally, we present a novel conceptual framework for the photoassociation of Cs$_2$ molecules through dual-laser illumination. By engineering light-induced potential energy surfaces  and leveraging the dependency of transition probabilities on laser phase, we have outlined a method for creating double-well-shaped potentials. These potentials are designed to easy transitions between long-range bound states and the molecule's electronic ground and vibrational states.

The implications of our study extend beyond the immediate results, offering a template for controlling molecular dynamics via sophisticated light manipulation, which could have far-reaching impacts in the field of quantum control and molecular physics.

\begin{acknowledgments}
We Acknowledge financial support by FONDECYT through grants 1220366 and 1211038 and by EU Horizon 2020 research and innovation programme under Maria-Sklodowska-Curie Grant Agreement 837028 (HYDROTRONICS Project) . L.E.F.F.T also Acknowledges the support of The Abdus Salam International Center for Theoretical Physics and Simons Foundation.  C.C. acknowledges support by Center for the Development of Nanosciences and Nanotechnology, CEDENNA AFB 220001. E.B. Acknowledges the support of ANID  Chile through the Doctoral National Scholarship N$^{\circ}\ 21192248$. Powered@NLHPC: This research was partially supported by the supercomputing infrastructure of the NLHPC (ECM-02). 
\end{acknowledgments}

\appendix
\appendix{\counterwithin{equation}{section}}
\appendix
\section{\label{Appendix:Quasiperiodic_driving} Quasiperiodic driving}

We want to find the Floquet modes through the eigenvalue equation \eqref{eq:HF_eigenvalue}
\begin{align}\label{eq_App:HF_eigenvalue}
    \left(\mathcal{H}(t)-i\hbar\frac{\partial}{\partial t}\right)\Ket{\phi_j(t)}=\varepsilon_j\Ket{\phi_j(t)}
\end{align}
To solve Eq.~\eqref{eq_App:HF_eigenvalue}, the Floquet modes and $\mathcal{H}_{\rm int}(t)$ are expanded in a double Fourier series
\begin{align}\label{eq:floquet_modes}
  \Ket{\phi_j(t)} &= \sum_{k_1=-\infty}^{\infty}\sum_{k_2=-\infty}^{\infty} \Ket{F_{j,k_1,k_2}}e^{ik_1\omega_1 t}e^{ik_2\omega_2t}\,,\\
  \mathcal{H}_{\rm{int}}(t) &= \sum_{k_1=-\infty}^{\infty}\sum_{k_2=-\infty}^{\infty} \mathcal{H}_{\rm int}^{[k_1,k_2]}e^{ik_1\omega_1 t}e^{ik_2\omega_2t}\,.
\end{align}

Replacing Eq.~\eqref{eq:floquet_modes} in Eq.~\eqref{eq_App:HF_eigenvalue} and introducing the vector notation $\vec k\cdot \vec\omega=k_1\omega_1+k_2\omega_2$ results in
\begin{align}\label{eq:fourier_expansion}
 \nonumber &\varepsilon_j\sum_{\vec k}\Ket{F_{j,\vec k}}e^{i\vec k\cdot\vec \omega t}-\sum_{\vec k}\hbar\vec k\cdot\vec\omega \Ket{F_{j,\vec k}}e^{i\vec k\cdot\vec\omega t} \\
  &= \sum_{\vec k} \mathcal{H}_e \Ket{F_{j,\vec k}} e^{i\vec k\cdot\vec\omega t}+\sum_{\vec k,\vec l}\mathcal{H}_{{\rm int}}^{\left[\,\vec l\:\right]}\Ket{F_{j,\vec k}} e^{i\left(\vec k+\vec l\:\right)\cdot \vec\omega t}\,.
\end{align}

To find an equation for the $\ket{F_{j,\vec k}}$ we must multiply Eq.~\eqref{eq:fourier_expansion} by $e^{-i\vec n\cdot\vec\omega t}$ and perform a quasiperiodic time-average~\cite{schilder2006}. This procedure will give rise to the Kronecker delta for multiperiodic Fourier series~\cite{osborne2018}
\begin{align}\label{eq:multiperiodic_delta}
  \delta_{\vec k,\vec n} &= \lim_{T\to\infty}\frac{1}{2T}\int\limits_{-T}^{T} e^{i\left(\vec k-\vec n\,\right)\cdot\vec\omega t} dt\,,
\end{align}
allowing us to finally arrive at the equation
\begin{align}\label{eq:F_jn1n2}
  \nonumber \varepsilon_j \Ket{F_{j,n_1,n_2}} &=\mathcal{H}_e \Ket{F_{j,n_1,n_2}} +\hbar(n_1\omega_1+n_2\omega_2)\Ket{F_{j,n_1,n_2}}\\
  &+\sum_{k_1,k_2}\mathcal{H}_{{\rm int}}^{[n_1-k_1,n_2-k_2]}\Ket{F_{j,k_1,k_2}}\,.
\end{align}

It is convenient to expand $\Ket{F_{j,n_1,n_2}}$ in the eigenstates~$\Ket{\zeta_\ell}$ of electronic Hamiltonian in the BO approximation ($\mathcal{H}_e\Ket{\zeta_\ell}=V_\ell\Ket{\zeta_\ell}$)
\begin{align}\label{eq:Fjn1n2_expansion}
    \Ket{F_{j,n_1,n_2}} &= \sum_{\ell} C_{n_1,n_2}^{\ell,j}\Ket{\zeta_\ell}\,. 
\end{align}

Using the expansion \eqref{eq:Fjn1n2_expansion} in Eq.~\eqref{eq:F_jn1n2} and projecting onto $\bra{\zeta_l}$, we obtain
\begin{align}\label{eq:C_ljn1n2}
  \nonumber \varepsilon_jC_{n_1,n_2}^{l,j} &=\left(V_l+\hbar n_1\omega_1+\hbar n_2\omega_2\right)C_{n_1,n_2}^{l,j}\\
  &+\sum_{\ell,k_1,k_2}C_{k_1,k_2}^{l,j}
  \left\langle \zeta_l\left|\mathcal{H}_{\rm int}^{[n_1-k_1,n_2-k_2]} \right|\zeta_\ell\right\rangle\,.
\end{align}

Eq.~\eqref{eq:C_ljn1n2} is valid provided that the ratio $\omega_1/\omega_2$ is irrational and may be reliable in the commensurate case. The latter situation is comprehensively discussed in reference \cite{poertner2020}.

Eq.~\eqref{eq:C_ljn1n2} can be  written  as
\begin{align}\label{eq:C_ljn1n2_ShirleyForm}
    \nonumber \varepsilon_j C_{n_1,n_2}^{l,j} &=\sum_{\ell,k_1,k_2}\left(H_{l\ell}^{[n_1-k_1,n_2-k_2]}\right.\\
     &\left.+\hbar(k_1\omega_1+k_2\omega_2)\vphantom{H_{l\ell}^{[n_1-k_1,n_2-k_2]}}\delta_{l\ell}\delta_{n_1k_1}\delta_{n_2k_2}\right)C_{k_1,k_2}^{\ell,j}\,,
\end{align}
where $H_{l\ell}^{[n_1-k_1,n_2-k_2]}$ is given by
\begin{align}
      \nonumber H_{l\ell}^{[n_1-k_1,n_2-k_2]} &=V_l\!\: \delta_{l\!\:\ell}\delta_{n_1k_1}\delta_{n_2k_2}\\
    &+ \left\langle \zeta_l\left|\mathcal{H}_{\rm int}^{[n_1-k_1,n_2-k_2]} \right|\zeta_\ell\right\rangle.
\end{align}

Eq.\eqref{eq:C_ljn1n2_ShirleyForm} is an extension of Shirley's formulation of Floquet Theory for periodic driving (see equation (8) in \cite{shirley1965}) and is equivalent to the Many-Mode Floquet Theory (MMFT) of Ho \textit{et al}. \cite{ho1983,ho1984}. The purpose behind the derivation of Eq.\eqref{eq:C_ljn1n2_ShirleyForm} through the expansion of Floquet modes in a double Fourier series and not directly lean on the Floquet Hamiltonian from MMFT is to  introduce the quasiperiodic time-average \eqref{eq:multiperiodic_delta}. This allow us to reveal the orthonormalization of Floquet modes,
\begin{align}
  \delta_{j^{\prime}j} &=\lim_{T\to\infty}\frac{1}{2T}\int\limits_{-T}^{T}\braket{\phi_{j^{\prime}}(t)|\phi_j(t)}dt=\left\langle \right\langle \phi_{j^{\prime}}\left.\right|\phi_j \left\rangle\right\rangle.
\end{align}

Thus, from all previous theoretical development, the $\ket{\phi_j(t)}$ can finally be expressed as
\begin{align}\label{eq:phi_double_fourier}
    \ket{\phi_j(t)} &=\sum_{l,n_1,n_2}C_{n_1,n_2}^{l,j} \Ket{\zeta_l}e^{in_1\omega_1 t}e^{in_2\omega_2 t}\,,
\end{align}
and their orthonormalization can be written as
\begin{align}\label{eq:DeltaKronecker_phi}
    \delta_{j^{\prime}j} &= \timebraket[-0.05cm]{\phi_{j^{\prime}}\!}{\phi_j\!}=\sum_{\ell,n_1,n_2}C_{n_1,n_2}^{\ell,j^{\prime}\!\; \ast} C_{n_1,n_2}^{\ell,j}\,.
\end{align}

\section{\label{Appendix:derivative_couplings} Nonadiabatic derivative couplings}

We start by expanding the total wavefunction using the Floquet modes as electronic basis
\begin{align}\label{eq_App:nuclear_floquet_ansatz}
    \Ket{\Psi(t)} &=\sum\limits_{j} \chi_j\Ket{\phi_j(t)}.
\end{align}

Replacing Eq.~\eqref{eq_App:nuclear_floquet_ansatz} in the time-dependent Schr\"odinger equation \eqref{eq:schrodinger}, we obtain
\begin{align}\label{eq:BornHuang_Kinetic+HF}
    \nonumber i\hbar\sum\limits_{j}\frac{\partial \chi_j}{\partial t}\Ket{\phi_j} &= \sum\limits_{j} \hat T\left(\chi_j\Ket{\phi_j}\right)\\
    &+\sum\limits_{j}\chi_j\left[\mathcal{H}(t)-i\hbar\frac{\partial}{\partial t}\right]\Ket{\phi_j}.
\end{align}

The Floquet operator \eqref{eq:HF_operator} can be recognized in the last term of the right-hand side of Eq.~\eqref{eq:BornHuang_Kinetic+HF}. Thereby, we have
\begin{align}\label{eq:chi_phi}
     i\hbar\sum\limits_{j}\frac{\partial \chi_j}{\partial t}\Ket{\phi_j} &= \sum\limits_{j} \hat T\left(\chi_j\Ket{\phi_j}\right)+\sum\limits_{j}\chi_j\varepsilon_j\Ket{\phi_j}.
\end{align}

In view of the fact that we are interested on the diatomic molecule Cs$_2$, the kinetic energy operator is just $-\hbar^2\nabla^2/(2\mu)$ ($\mu$ denote the reduced mass). Therefore,  Eq.~\eqref{eq:chi_phi} can be rewritten as
\begin{align}\label{eq:full_chi_phi}
\nonumber i\hbar\sum\limits_{j}\frac{\partial \chi_j}{\partial t}\Ket{\phi_j} &= -\frac{\hbar^2}{2\mu} \sum\limits_{j}\left(\nabla^2\chi_j\right) \Ket{\phi_j}+\sum\limits_{j}\chi_j\varepsilon_j\Ket{\phi_j}\\
 &-\frac{\hbar^2}{2\mu} \sum\limits_{j}\left(\nabla^2\Ket{\phi_j}+2\nabla\ket{\phi_j}\cdot\nabla \right)\chi_j\,.
\end{align}

We cannot take the quasiperiodic average \eqref{eq:multiperiodic_delta} directly over Eq.~\eqref{eq:full_chi_phi} to derive an equation of motion for $\chi_j$. Instead, we must proceed by expressing $\ket{\phi_j}$ in its double Fourier series form \eqref{eq:phi_double_fourier}. By inserting Eq.~\eqref{eq:phi_double_fourier} in Eq.~\eqref{eq:full_chi_phi}, the harmonic factor $e^{i\vec n\cdot\vec\omega t}$ ($\vec n\cdot\vec \omega=n_1\omega_1+n_2\omega_2$) can be factorized in every term of Eq.~\eqref{eq:full_chi_phi}. Hence, the resulting equation is
\begin{align}\label{eq:nabla_chi_zeta}
    \nonumber 0 &= \sum\limits_{\vec n}e^{i\vec n\cdot\vec\omega t}\left[\sum\limits_{j,\,\ell}\left( -i\hbar\frac{\partial\chi_j}{\partial t}C_{\vec n}^{\ell,j}\Ket{\zeta_\ell}\right.\right.\\
    \nonumber &-\frac{\hbar^2}{2\mu}\nabla^2\chi_j\,C_{\vec n}^{\ell,j}\Ket{\zeta_\ell}-\frac{\hbar^2}{\mu}\nabla\left\{C_{\vec n}^{\ell,j}\Ket{\zeta_\ell}\right\}\cdot\nabla\chi_j\\
     &-\frac{\hbar^2}{2\mu}\nabla^2\left\{C_{\vec n}^{\ell,j}\Ket{\zeta_\ell}\right\}\chi_j\,+\chi_j\,\varepsilon_j\,C_{\vec n}^{\ell,j}\Ket{\zeta_\ell}\left.\vphantom{\frac{\partial\chi_j}{\partial t}}\right)\left.\vphantom{\sum\limits_{j,\,\ell}}\right],
\end{align}
from which the square bracket must be zero at any time, since $e^{i\vec n\cdot\vec\omega t}\neq 0$. The next step is to operate with $\nabla$ in Eq.~\eqref{eq:nabla_chi_zeta} over the $C_{\vec n}^{\ell,j}$ coefficients and $\Ket{\zeta_\ell}$. After the algebra related to $\nabla$,  we need to project onto $\bra{\zeta_l}$, obtaining the equation
\begin{align}\label{eq:chi_Cnlj}
    \nonumber i\hbar\sum\limits_{j}\frac{\partial \chi_j}{\partial t} C_{\vec n}^{l,j} &= -\frac{\hbar^2}{2\mu}\sum\limits_{j}\nabla^2\chi_j\!\:C_{\vec n}^{l,j}+\sum\limits_{j}\chi_j\!\:\varepsilon_j\!\:C_{\vec n}^{l,j} \\
    \nonumber &-\frac{\hbar^2}{2\mu}\sum\limits_{j}\chi_j\nabla^2 C_{\vec n}^{l,j}-\frac{\hbar^2}{\mu}\sum\limits_{j}\nabla\chi_j\cdot\nabla C_{\vec n}^{l,j}\\
    \nonumber&-\frac{\hbar^2}{2\mu}\sum\limits_{j,\!\: \ell}C_{\vec n}^{\ell,j}\;\chi_j\Braket{\zeta_l|\nabla^2\zeta_\ell}\\
    \nonumber &-\frac{\hbar^2}{\mu}\sum\limits_{j,\!\: \ell}C_{\vec n}^{\ell,j}\;\nabla\chi_j\cdot\Braket{\zeta_l|\nabla\zeta_\ell}\\
    &-\frac{\hbar^2}{\mu}\sum\limits_{j,\!\: \ell}\;\chi_j\nabla C_{\vec n}^{\ell,j}\cdot\Braket{\zeta_l|\nabla\zeta_\ell}
\end{align}

To attain the equation for the nuclear wavefunctions, we need to multiply Eq.~\eqref{eq:chi_Cnlj} by $C_{\vec n}^{l,k\!\:\ast}$ and sum over all $\vec n$ and $l$. In this way we obtain  
\begin{align}\label{eq:chi_despejado}
    \nonumber i\hbar\frac{\partial \chi_k}{\partial t} &=-\frac{\hbar^2}{2\mu} \nabla^2\chi_k+\varepsilon_k\chi_k-\frac{\hbar^2}{2\mu}\sum\limits_{j,\!\: l,\vec n}\chi_j\!\: C_{\vec n}^{l,k\!\:\ast}\nabla^2C_{\vec n}^{l,j}\\
    \nonumber&-\frac{\hbar^2}{2\mu}\sum\limits_{j,\!\: l,\!\:\ell,\vec n}\chi_j\;C_{\vec n}^{l,k\!\:\ast}\Braket{\zeta_l|\nabla^2\zeta_\ell}C_{\vec n}^{\ell,j}\\
    \nonumber &-\frac{\hbar^2}{\mu}\sum\limits_{j,\!\: l,\!\:\ell,\vec n}\;\chi_j C_{\vec n}^{l,k\!\:\ast}\Braket{\zeta_l|\nabla\zeta_\ell}\cdot\nabla C_{\vec n}^{\ell,j},\\
    \nonumber &-\frac{\hbar^2}{\mu}\sum\limits_{j,\!\: l,\vec n}\nabla\chi_j\cdot C_{\vec n}^{l,k\!\:\ast}\nabla C_{\vec n}^{l,j}\\
     &-\frac{\hbar^2}{\mu}\sum\limits_{j,\!\: l,\!\:\ell,\vec n}\!\:\nabla\chi_j\cdot C_{\vec n}^{l,k\!\:\ast}\Braket{\zeta_l|\nabla\zeta_\ell}C_{\vec n}^{\ell,j}.
\end{align}

By calculating the time-average over the derivatives of Floquet modes, namely
\begin{align}
\nonumber \timebraket[-0.08cm]{\!\!\phi_k\!}{\!\nabla^2\phi_j\!\!} &= \left( \sum\limits_{l,\vec n^{\prime}} C_{\vec n^{\prime}}^{l,k\!\:\ast} \Bra{\zeta_{l}}\right)\nabla^2\left(\sum\limits_{\ell,\vec n} C_{\vec n}^{\ell,j} \Ket{\zeta_\ell}\right)\delta_{\vec n^{\prime}\vec n}\,,\\
    \nonumber &= \sum\limits_{l,\vec n} C_{\vec n}^{l,k\!\:\ast}\nabla^2 C_{\vec n}^{l,j}+\sum\limits_{l,\ell,\vec n}C_{\vec n}^{l,k\!\:\ast} \Braket{\zeta_l|\nabla^2\zeta_\ell}C_{\vec n}^{\ell,j}\\
    &+2\sum\limits_{l,\ell,\vec n}C_{\vec n}^{l,k\!\:\ast} \Braket{\zeta_l|\nabla\zeta_\ell}\cdot\nabla C_{\vec n}^{\ell,j},\\
    \nonumber \timebraket[-0.05cm]{\!\!\:\phi_k\!}{\!\nabla\phi_j\!\!} &= \left( \sum\limits_{l,\vec n^{\prime}} C_{\vec n^{\prime}}^{l,k\!\:\ast} \Bra{\zeta_{l}}\right)\nabla\left(\sum\limits_{\ell,\vec n} C_{\vec n}^{\ell,j} \Ket{\zeta_\ell}\right)\delta_{\vec n^{\prime}\vec n}\\
    \nonumber &= \sum\limits_{l,\vec n} C_{\vec n}^{l,k\!\:\ast}\nabla C_{\vec n}^{l,j}+\sum\limits_{l,\ell,\vec n}C_{\vec n}^{l,k\!\:\ast} \Braket{\zeta_l|\nabla\zeta_\ell}C_{\vec n}^{\ell,j},
\end{align}
the Eq.\eqref{eq:chi_despejado} can be represented in the following compacted form
\begin{align}
    \nonumber i\hbar\frac{\partial \chi_k}{\partial t} &=-\frac{\hbar^2}{2\mu} \nabla^2\chi_k+\varepsilon_k\chi_k-\frac{\hbar^2}{2\mu}\sum\limits_{j}\chi_j\timebraket[-0.08cm]{\!\!\phi_j\!}{\!\nabla^2\phi_k\!\!}\\
     &-\frac{\hbar^2}{\mu}\sum\limits_{j}\nabla\chi_j\cdot\timebraket[-0.05cm]{\!\!\:\phi_j\!}{\!\nabla\phi_k\!\!}.
\end{align}

\bibliographystyle{apsrev4-1_title}
\bibliography{references}

\end{document}